# MILLIMETER WAVE THIN-FILM BULK ACOUSTIC RESONATOR IN SPUTTERED SCANDIUM ALUMINUM NITRIDE USING PLATINUM ELECTRODES

*Sinwoo Cho[1], Omar Barrera[1], Pietro Simeoni[2], Ellie Y. Wang[1], Jack Kramer[1], Vakhtang Chulukhadze[1], Joshua Campbell[1], Matteo Rinaldi[2], and Ruochen Lu[1]*
[1]The University of Texas at Austin, US, [2]Northeastern University, US

## ABSTRACT

This work describes sputtered scandium aluminum nitride (ScAlN) thin-film bulk acoustic resonators (FBAR) at millimeter wave (mmWave) with high quality factor ($Q$) using platinum (Pt) electrodes. FBARs with combinations of Pt and aluminum (Al) electrodes, i.e., Al top – Al bottom, Pt top – Al bottom, Al top – Pt bottom, and Pt top – Pt bottom, are built to study the impact of electrodes. The FBAR with Pt top – Pt bottom achieves electromechanical coupling ($k^2$) of 1.8% and $Q$ of 94 for third-order symmetric (S3) mode at 61.6 GHz, confirming FBAR can achieve a $Q$ approaching 100 with optimized fabrication and acoustic design.

## INTRODUCTION

Radio frequency (RF) piezoelectric devices are key technology for sub-6 GHz RF filtering resolution [1]–[4]. Electrically coupled piezoelectric resonators transform electromagnetic (EM) waves into mechanical vibrations at mechanical resonances [5], [6]. Such conversion presents two advantages over EM equivalents: efficient frequency selectivity and miniature footprints [2]. One commercially successful technology is thin-film bulk acoustic wave resonators (FBAR) in sputtered scandium aluminum nitride (ScAlN) and aluminum nitride (AlN), thanks to high electromechanical coupling ($k^2$), quality factor ($Q$), and well-established CMOS compatible microfabrication [2], [3], [7]. As wireless communication advances into the millimeter-wave (mmWave) bands exceeding 30 GHz, there is a chance to scale the frequency of ScAlN/AlN devices while preserving high performance [3].

However, previous research indicates the difficulty of scaling ScAlN/AlN FBARs, as demonstrated by the presentation of moderate figures of merit (FoM, $k^2 \cdot Q$) at mmWave bands (Fig. 7). First, mmWave FBAR requires a film stack of sub-100 nm thick piezoelectric and sub-50 nm electrode layers. Keeping good crystalline during sputtering is challenging due to surface roughness and oxygen level in intermediate boundaries [8]–[10]. Secondly, few studies analyze the impact of different electrode materials on $k^2$ and $Q$ at mmWave [11], [12]. The effects of electrodes are essential, especially for resonators operating in higher-order thickness modes (e.g., third-order symmetric, S3) where a significant portion of energy lies in the electrodes [8], [9]. Thirdly, increased electrical resistance and inductance in thinner electrodes and interconnects are not well studied [10]. Metal resistance and stress rise sharply with unoptimized sputtering gas

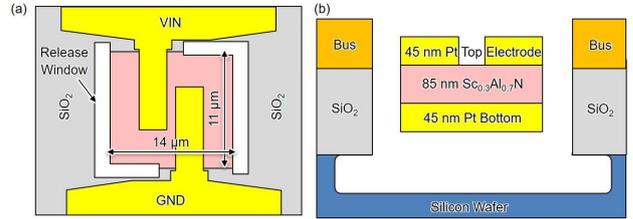

*Fig. 1 (a) Top and (b) cross-sectional view of ScAlN mmWave FBAR.*

variety and deposition temperature [13]. Lastly, because the capacitance density in thinner films grows noticeably, new acoustic designs and microfabrication techniques will be required for 50 Ω systems, necessitating a large reduction in the resonator's lateral dimensions. All of the above challenges require innovative design and fabrication toward mmWave FBARs.

Recently, studies on mmWave AlN/ScAlN FBARs have focused on piezoelectric material and acoustic design, toward demonstrating the possibility of achieving resonator $Q$ beyond 100 while keeping high $k^2$ at mmWave. Better high-quality crystallinity ScAlN/AlN films have been created via molecular beam epitaxy (MBE) and metal-organic chemical vapor deposition (MOCVD) [14]. Despite the great film quality, the integration issues with bottom electrodes and sacrificial layers lead to moderate performance increases [15]–[17]. Alternatively, innovative acoustic designs, including overmoded resonators in AlN, have been examined and have shown superior performance because of a greater volume-to-surface ratio, which reduces surface-induced loss [15], [18]–[21]. Nevertheless, The devices' performance is still behind their equivalents at lower frequencies [22]. The performance bottleneck may lie outside the piezoelectric layer. The electrodes might play an important role in mmWave FBAR performance, but direct investigation of their impact is insufficient.

In this work, we report the influence of metal electrodes (Pt and Al) on sputtered ScAlN FBARs at mmWave by comparing a group of four film stacks with a combination of Pt and Al as top and bottom electrodes. Third-order symmetric (S3) mode devices at 50 GHz and above are implemented with sputtered $Sc_{0.3}Al_{0.7}N$. The demonstrated FBAR with Pt top and bottom electrodes achieved $k^2$ of 4.0% and $Q$ of 116 for the first-order symmetric (S1) mode at 13.7 GHz, and $k^2$ of 1.8% and $Q$ of 94 for S3 at 61.6 GHz. Through these results, we confirmed that even in the frequency band of approximately 60 GHz, FBAR can achieve a $Q$ factor approaching 100 with optimized fabrication and acoustic/EM design. Further development calls for better-quality film stacks in both piezoelectric and metallic layers.

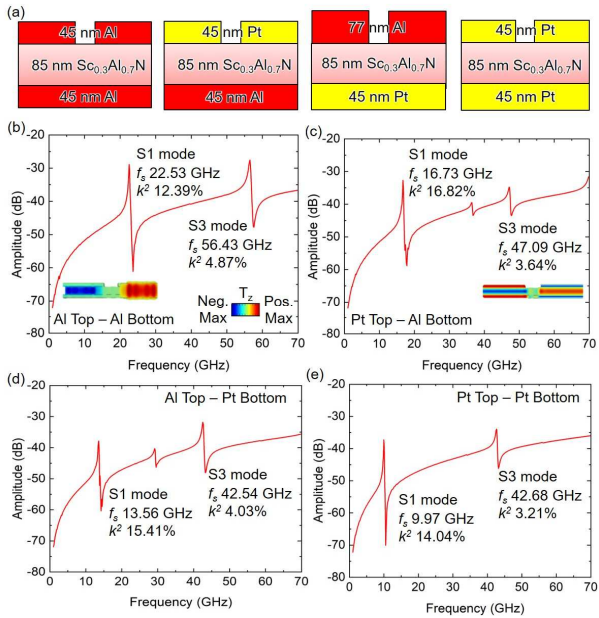

*Fig. 2 (a) Four resonator stacks with different metals. Wideband admittance of (b) Al top – Al bottom (c) Pt top – Al bottom (d) Al top – Pt bottom (e) Pt top – Pt bottom.*

## DESIGN AND SIMULATION

The top and cross-sectional views of the FBAR are displayed in Fig. 1 (a)-(b). The film stack comprises a bottom floating electrode, top electrodes with ground and signal traces, and 85 nm thick $Sc_{0.3}Al_{0.7}N$ sandwiched between 45 nm thick Pt top and bottom electrodes. This stack thickness enables S3 mode in mmWave bands over 30 GHz. As a result of the high capacitance density of the thin film, the lateral dimensions of the resonant body are minimized to 14 by 11 μm. 400 nm bulines are used to reduce routing resistance. The lateral $SiO_2$ layer surrounding the resonant body dramatically decreases the feedthrough-induced parasitic resistance and capacitance.

The proposed 4 different FBAR stacks for analyzing impacts of the top and bottom electrodes [Fig. 2 (a)] are simulated using COMSOL finite element analysis (FEA) with a mechanical $Q$ value of 50, estimated from prior measurement [8], [9]. The S3 mode shape in Fig. 2 (c) verifies that the stack maximizes $k^2$ for S3, as the stress node lies in the metal-ScAlN interfaces. The Al top – Al bottom resonator shows S1 at 22.53 GHz with $k^2$ of 12.39%, and S3 at 56.43 GHz with $k^2$ of 4.87% [Fig. 2 (b)]. The Pt top – Al bottom resonator displays S1 at 22.53 GHz with $k^2$ of 12.39% and the S3 mode at 56.43 GHz with $k^2$ of 4.87% [Fig. 2 (c)]. The Al top – Pt bottom resonator shows S1 at 13.56 GHz with $k^2$ of 15.41%, and S3 at 42.54 GHz with $k^2$ of 4.03% [Fig. 2 (d)]. The Pt top – Pt bottom resonator displays S1 at 9.97 GHz with $k^2$ of 14.04%, and S3 at 42.68 GHz with $k^2$ of 3.21% [Fig. 2(e)]. Due to the difference in the metal properties of Pt and Al, second-order antisymmetric (A2) modes are observed in Fig 2 (c)-(d). $k^2$ of S3 mode was larger in designs using Al electrodes than Pt electrodes due to less density in Al. Thus, it is expected that the selection of the appropriate metal for FBAR predicts that device performance can be maximized.

## MATERIAL ANALYSIS & FABRICATION

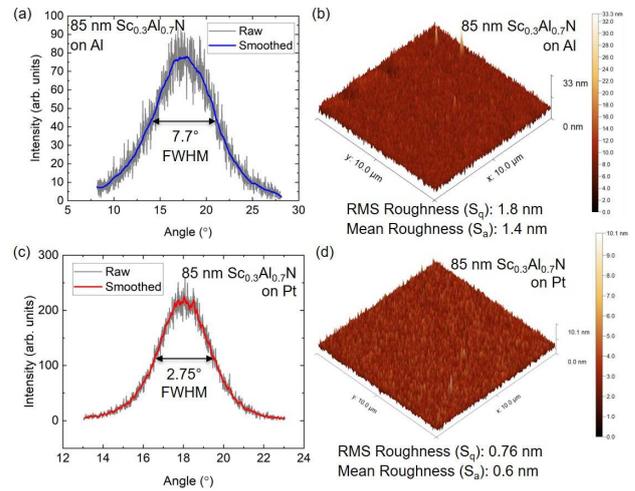

*Fig. 3 Analysis of ScAlN on Al by (a) XRD (b) AFM. Analysis of ScAlN on Pt by (c) XRD (d) AFM.*

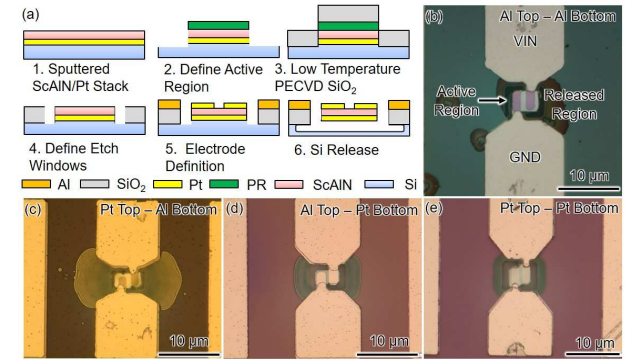

*Fig. 4 (a) FBAR fabrication flow chart and its device images of (b) Al top – Al bottom (c) Pt top – Al bottom (d) Al top – Pt bottom (e) Pt top – Pt bottom.*

The fabrication process begins with depositing a layer of 45 nm thick Al or Pt and an 85 nm layer of ScAlN onto a high-resistivity (> 10,000 Ω·cm) Si <100> wafer. The Evatec Clusterline 200 sputtering tool is used for this deposition, which keeps an uninterrupted vacuum environment. X-ray diffraction (XRD) is the first step in the quantitative material analysis for each stack [Fig. 3 (a) and (c)]. The $Sc_{0.3}Al_{0.7}N$ full width at half maximum (FWHM) of the rocking curve on the bottom Al electrode is 7.7°, indicating that the sputtered thin film has non-ideal crystal quality [23]. The FWHM of the counterpart based on the Pt rocking curve is 2.75°, suggesting better crystal quality. Fig. 3 (b) and (d) show the surface roughness of the sputtered ScAlN film from the atomic force microscopy (AFM). Sputtered ScAlN film onto Pt shows better surface uniformity than the Al counterpart [24].

Fig. 4 (a) illustrates the fabrication process. The first step involves etching the layers of ScAlN, Al, and Si surrounding the active region using the AJA Ion Mill process. Following this, a layer of $SiO_2$ is deposited on these etched regions by low-temperature (100 °C) plasma-enhanced chemical vapor deposition (PECVD) [25]. This $SiO_2$ layer guarantees electrical isolation and protects against the top electrode disconnected from height difference. Next, the AJA Ion Mill is used to establish and etch release windows [26]. After that, Al bus lines and pads are deposited together with top electrodes using a KJL e-beam evaporator. The structural release is then confirmed by axenon difluoride (XeF$_2$) Si isotropic etching procedure. Optical depictions of fabricated FBAR combinations are presented in Fig. 4 (b)-(e). FBAR devices have similar designs and dimensions.

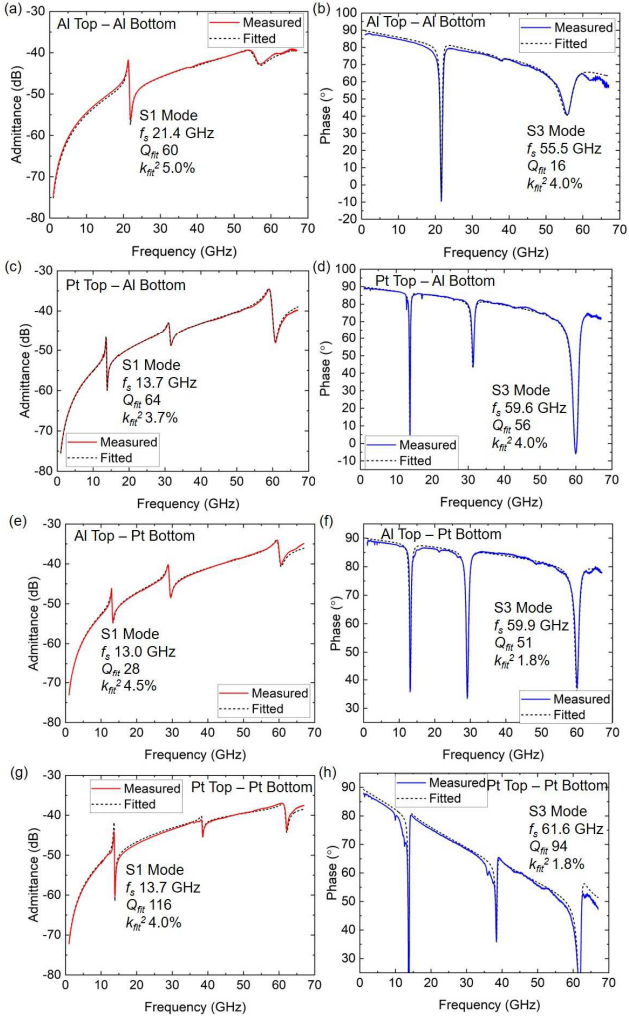

*Fig. 5 Measured admittance in amplitude and phase (a)-(b) Al top – Al bottom (c)-(d) Pt top – Al bottom (e)-(f) Al top – Pt bottom (g)-(h) Pt top – Pt bottom.*

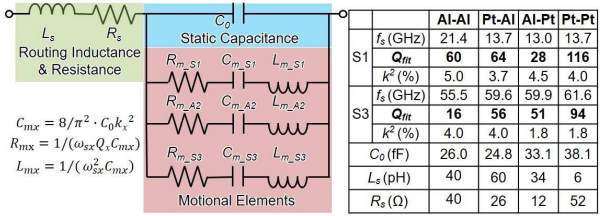

*Fig. 6 Modified mBVD model and extracted key resonator parameters for four cases.*

## MEASUREMENT AND DISCUSSION

Resonators are measured with a Keysight vector network analyzer (VNA) by a two-port setup within room temperature air, at −15 dBm [27]. Fig. 5 (a)–(h) displays the obtained admittance amplitude and phase, demonstrating the existence of the S1 and S3 modes. The A2 mode, caused by the difference in thickness between the upper and lower electrodes, generates a minor resonance between S1 and S3. Measured wideband admittance of each FBAR in amplitude and phase, including Al top-Al bottom, Pt top-Al bottom, Al top-Pt bottom, and Pt top-Pt bottom, is described in Fig. (a)-(b), Fig. (c)-(d), Fig. (e)-(f), and Fig. (g)-(h) in order. A modified Butterworth Van Dyke (mBVD) model is exploited to extract each resonator performance in Fig. 6, putting series routing resistance ($R_s$) and inductance ($L_s$) to capture the electromagnetic (EM) effects. Before adding the motional elements for extracting $Q$ and $k^2$ in Fig. 6, the EM parameters, i.e., $R_s$, $L_s$, and static capacitance $C_0$, are first fitted from the admittance amplitude and phase [Fig. 5 (a)-(h)]. The fitted curves are plotted in Fig. 7 (a)-(h). The Pt top – Al bottom FBAR in Fig.5 (c)-(d) achieves $k^2$ of 3.7% and $Q_{fit}$ of 64 for S1 mode, along with $k^2$ of 4.0% and $Q_{fit}$ of 56 for S3 mode, leading to FoM of 2.37 and 2.24. Its high FoM at S3 mode shows the possibility of proper usage of metal electrodes might achieve good performance of resonator with enhanced $Q$ and $k^2$ at mmWave. The Pt top – Pt bottom FBAR in Fig.5 (g)-(h) achieves $k^2$ of 4.0% and $Q_{fit}$ of 116 for S1 mode, along with $k^2$ of 1.8% and $Q_{fit}$ of 94 for S3 mode. Its $Q_{fit}$ shows the possibility of achieving resonator $Q$ beyond 100 while keeping high $k^2$ at mmWave.

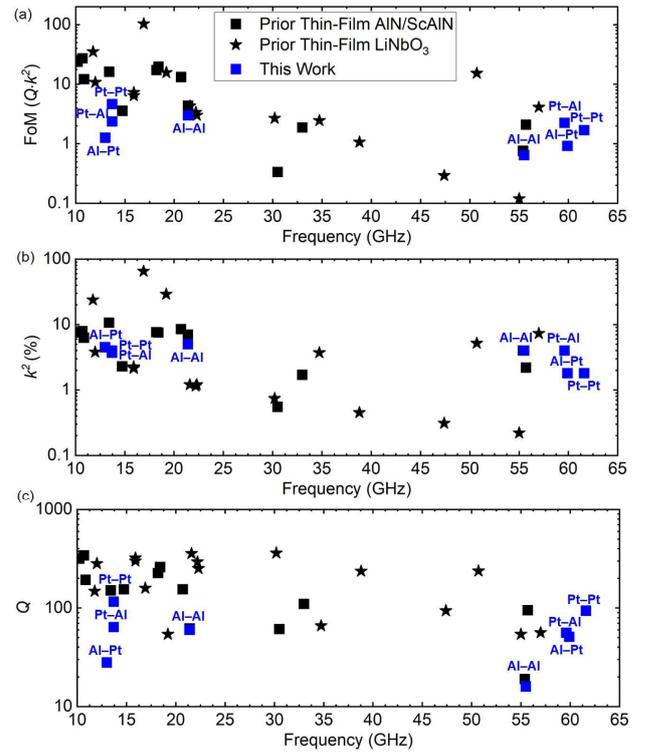

*Fig. 7 Survey of reported resonators above 10 GHz (a) FoM (b) $k^2$ (c) Q.*

Fig. 7 shows studies of FoM, $k^2$, and $Q$ for reported resonators above 10 GHz, including the AlN/ScAlN [18]–[21] and lithium niobate (LiNbO$_3$) demonstrations [28]–[30], compared with the state-of-the-art (SoA). Our study exhibits comparable FoM to previous ScAlN/AlN works for the 13.7 GHz, while around 59.6 GHz devices show higher FoM than previous ScAlN/AlN works. Nevertheless, the FoM is less than that of transferred thin-film LiNbO$_3$-based mmWave resonators with significantly superior film quality (<100 arcsec FWHM) [29], suggesting that greater crystalline quality from improved deposition techniques is necessary for the continued development of ScAlN mmWave resonators.

## CONCLUSION

We reported ScAlN FBARs with a high $Q$ of 94 using a Pt electrode at 61.6 GHz. Via studying FBARs with Al and Pt top and bottom electrodes, we notice Pt devices show higher $Q$ at the cost of $k^2$ for S3 mode. These results confirmed that even in the frequency band of approximately 60 GHz, ScAlN FBAR can achieve a Q factor approaching 100 with optimized fabrication and acoustic/EM design. Material-level analysis and device-level performance show that the bottleneck for better performance enhancement lies in better stack quality in piezoelectric and metallic layers.

## ACKNOWLEDGEMENTS
The authors would like to thank the funding support from the DARPA COFFEE program and Dr. Ben Griffin for the helpful discussion.

## REFERENCES

[1] S. Gong *et al.*, "Microwave Acoustic Devices: Recent Advances and Outlook," *IEEE Microw. Mag.*, vol. 1, no. 2, 2021.

[2] R. Ruby, "A snapshot in time: The future in filters for cell phones," *IEEE Microw Mag*, vol. 16, no. 7, 2015.

[3] A. Hagelauer *et al.*, "From Microwave Acoustic Filters to Millimeter-Wave Operation and New Applications," *IEEE Microw. Mag.*, vol. 3, no. 1, 2022.

[4] R. Lu *et al.*, "RF acoustic microsystems based on suspended lithium niobate thin films: Advances and outlook," *Journal of Microelectromechanical Systems*, vol. 31, no. 11, 2021.

[5] L. Gao, *et al.*, "Wideband Hybrid Monolithic Lithium Niobate Acoustic Filter in the K-Band," *IEEE TUFFC*, vol. 68, no. 4, 2021.

[6] R. Su *et al.*, "Lithium Niobate Thin Film Based A1Mode Resonators with Frequency up to 16 GHz and Electromechanical Coupling Factor Near 35%," in *IEEE MEMS*, 2023.

[7] G. Piazza *et al.*, "Piezoelectric aluminum nitride vibrating contour-mode MEMS resonators," *Journal of Microelectromechanical Systems*, vol. 15, no. 6, 2006.

[8] S. Cho *et al.*, "Millimeter Wave Thin-Film Bulk Acoustic Resonator in Sputtered Scandium Aluminum Nitride," *Journal of Microelectromechanical Systems*, pp. 1–4, 2023.

[9] S. Cho *et al.*, "55.4 GHz Bulk Acoustic Resonator in Thin-Film Scandium Aluminum Nitride," in *2023 IEEE IUS*, 2023

[10] K. Hashimoto, *RF bulk acoustic wave filters for communications*, vol. 66. 2009.

[11] G. F. Iriarte *et al.*, "Synthesis of C-axis oriented AlN thin films on metal layers: Al, Mo, Ti, TiN and Ni.," *IEEE IUS*, 2002.

[12] Z. Schaffer *et al.*, "Examination of Phonon Dissipation in 33 GHz Overmoded Bulk Acoustic Resonators," in *IEEE IUS*, 2022.

[13] J. A. Thornton *et al.*, "Internal stresses in metallic films deposited by cylindrical magnetron sputtering," *Thin Solid Films*, vol. 64, no. 1, 1979.

[14] M. T. Hardy *et al.*, "Nucleation control of high crystal quality heteroepitaxial $Sc_{0.4}Al_{0.6}N$ grown by molecular beam epitaxy," *J Appl Phys*, vol. 134, no. 10, 2023.

[15] S. Nam *et al.*, "A mm-Wave Trilayer AlN/ScAlN/AlN Higher Order Mode FBAR," *IEEE Microw. Wirel. Compon. Lett.*, vol. 33, no. 6, 2023.

[16] W. Zhao *et al.*, "X-band epi-BAW resonators," *J Appl Phys*, vol. 132, no. 2, 2022.

[17] M. Park *et al.*, "Epitaxial Aluminum Scandium Nitride Super High Frequency Acoustic Resonators," *Journal of Microelectromechanical Systems*, vol. 29, no. 4, 2020.

[18] Z. Schaffer *et al.*, "33 GHz Overmoded Bulk Acoustic Resonator," *IEEE Microw. Wirel. Compon. Lett.*, vol. 32, no. 6, 2022.

[19] R. Vetury *et al.*, "A Manufacturable AlScN Periodically Polarized Piezoelectric Film Bulk Acoustic Wave Resonator (AlScN P3F BAW) Operating in Overtone Mode at X and Ku Band," *IEEE IMS*, 2023, pp. 891–894.

[20] Izhar *et al.*, "A K-Band Bulk Acoustic Wave Resonator Using Periodically Poled $Al_{0.72}Sc_{0.28}N$," *IEEE Electron Device Lett.*, 2023.

[21] Z. Schaffer *et al.*, "A Solidly Mounted 55 GHz Overmoded Bulk Acoustic Resonator," in *2023 IEEE IUS*, 2023

[22] Y. Qu *et al.*, "Aluminum Nitride Based Film Bulk Acoustic Resonator With Anchor Column Structure," *Journal of Microelectromechanical Systems*, vol. 32, no. 2, 2023.

[23] X. Yi *et al.*, "High-Quality Film Bulk Acoustic Resonators Fabricated on AlN Films Grown by a New Two-Step Method," *IEEE Electron Device Letters*, vol. 43, no. 6, pp. 942–945, Jun. 2022.

[24] G. Giribaldi *et al.*, "High-Crystallinity 30% Scaln Enabling High Figure of Merit X-Band Microacoustic Resonators for Mid-Band 6G," in *IEEE MEMS*, 2023.

[25] C. Martinet *et al.*, "Deposition of SiO2 and TiO2 thin films by plasma enhanced chemical vapor deposition for antireflection coating," *J Non Cryst Solids*, vol. 216, 1997.

[26] V. Chulukhadze *et al.*, "Frequency Scaling Millimeter Wave Acoustic Resonators using Ion Beam Trimmed Lithium Niobate," *IFCS, 2023*.

[27] S. Cho *et al.*, "Analysis of 5-10 GHz Higher-Order Lamb Acoustic Waves in Thin-Film Scandium Aluminum Nitride," *IFCS-EFTF*, 2023.

[28] J. Kramer *et al.*, "57 GHz Acoustic Resonator with k2 of 7.3 % and Q of 56 in Thin-Film Lithium Niobate," in *2022 IEDM*, 2022

[29] J. Kramer *et al.*, "Thin-Film Lithium Niobate Acoustic Resonator with High Q of 237 and k2 of 5.1% at 50.74 GHz," in *IFCS-EFTF,* 2023

[30] J. Kramer *et al.*, "Trilayer Periodically Poled Piezoelectric Film Lithium Niobate Resonator," in *2023 IEEE IUS*, 2023

**CONTACT**

sinwoocho@utexas.edu